\documentclass[prl, twocolumn, showpacs, amsmath]{revtex4}
\usepackage{dcolumn}% Align table columns on decimal point
\usepackage{bm}% bold math
\usepackage{graphicx}% Include figure files
\usepackage{color}
\DeclareGraphicsExtensions{. jpg,. pdf, . mps, . png, . eps, . ps, . EPS}

\begin{document}
\def\be{\begin{equation}}
\def\ee{\end{equation}}

\def\bc{\begin{center}} 
\def\ec{\end{center}}
\def\bea{\begin{eqnarray}}
\def\eea{\end{eqnarray}}
\newcommand{\avg}[1]{\langle{#1}\rangle}
\newcommand{\Avg}[1]{\left\langle{#1}\right\rangle}

\title{Superconductor-Insulator Transition on Annealed Complex Networks }

\author{Ginestra Bianconi}

\affiliation{Department of Physics, Northeastern University, Boston, 
Massachusetts 02115 USA}
\begin{abstract} 
Cuprates show multiphase and multiscale complexity that has hindered physicists search for the mechanism of high $T_c$ for many years. Recently the interest has been addressed to a possible optimum inhomogeneity of dopants, defects and interstitials, and the structural scale invariance of dopants detected by scanning micro-x-ray diffraction has been reported to promote the critical temperature. In order to shed light on critical phenomena on granular materials, here we propose a stylized model capturing the essential characteristics of the superconducting-insulator transition of a highly dynamical, heterogeneous granular material: the Random Transverse Ising Model (RTIM) on Annealed Complex Network.
We show that when the networks encode for high heterogeneity of the expected degrees described by a power law distribution, the critical temperature for the onset of the superconducting  phase diverges to infinity as the power-law exponent $\gamma$ of the expected degree distribution is less than $3$, i.e. $\gamma<3$.
Moreover we investigate the case in which  the critical state of the electronic background is triggered by an external parameter $g$ that determines an exponential cutoff in the power law expected degree distribution characterized by an exponent $\gamma$.
We find that for $g=g_c$ the critical temperature for the superconducting-insulator transition has a maximum if $\gamma>3$ and diverges if $\gamma<3$.
\end{abstract}

\maketitle

Large attention  has been recently addressed to the effects that different topological properties may induce on the behavior of equilibrium and non-equilibrium processes defined on networks \cite{crit,Dynamics}. 
Heterogeneous degree distributions, small world and spectral properties, in particular, have been recognized as responsible of novel types of phase transitions and universality classes in classical processes \cite{crit,Dynamics,ising,Vespignani,Synchr}.
For instance, scale-free networks present a complex critical behavior of the Ising model, of percolation and of spreading processes, that 
explicitly depends on the exponent of the power-law in the degree distributions  \cite{crit,Dynamics,Vespignani}. On the other hand, the existence of non trivial spectral properties is crucial for  the stability of synchronization processes and $O(n)$ models \cite{Synchr}.
Nevertheless  there are no available results for   quantum critical phenomena \cite{sachdev} defined on  complex networks.

The aim of this paper is to investigate the role topological aspect of the network on the Random Transverse Ising Model .
In the last years there has been increasing interest in the understanding of the Random Transverse Ising Model  \cite{MaFisher,QCM,IoffeMezard1,IoffeMezard2}. This model that includes the disorder of onsite energies, has been used to characterize the insulator-superconducting phase transition in granular disordered superconductors \cite{IoffeMezard1,IoffeMezard2}. In each grain of granular materials the superconducting order parameter is well defined and the grains are coupled to each other by the  pair transfer term, so the physics is similar to the superconductivity in Josephson junction arrays \cite{Fazio}.
The new results on a  quenched Cayley tree have found the complex phase diagram of   the model by  the use  the quantum cavity method \cite{IoffeMezard1,IoffeMezard2,QCM}. 
The solution of the model is  able to  reproduce correctly the most important experimental features on disordered granular films: direct superconductor-insulator transition, activated behavior close to the quantum critical point \cite{Goldman} in the insulating phase, strong dependence of the activation energy near the quantum critical point and the order parameter of the superconducting phase variations from site to site \cite{Sacepe}. 

The cuprate superconductors that keep the record of the highest critical temperature (160 K)  of the phase diagram for all cuprate perovskite families \cite{stripe}
show a myriad of mysterious 'quantum matter' phenomena making these systems among the most intriguing puzzles in modern physics. In these doped Mott insulators the electrons form a poorly understood, highly collective quantum states. The strange electron matter of the cuprates strongly diverges from the standard weakly interacting quantum gas of conventional metals and superconductors. Here the electron matter is in a 'quantum-critical state', where the electrons form collective patterns that look the same regardless of scale. This scale free topology is  present both in space and in time scales,  because the electrons are in a state of perpetual quantum motion \cite{sachdev,Zaanen2004,Zaanen2009}. A key particular feature of the anomalous electronic matter of cuprates is the heterogeneity of the interactions in different spots of the $k$ space, going from weakly interacting electrons the so called "nodal points" to strongly interacting electrons at the "antinodal points" \cite{lanzara}. These materials show multiphase complexity (structural, magnetic and electronic) that has hindered physicists search for the mechanism of high $T_c$ for many years since the observed phenomenology depends on the time and spatial sensitivity of the experimental probes. 
The heterogeneity of cuprate superconductors has been proposed to be an essential feature of high Tc mechanism  \cite{muller,Dagotto1} and it has been proposed that above $T_c$ the electronic structure in the normal state is made of incoherent superconducting grains \cite{muller,Kresin1,Kresin2}. In fact cuprates show a similar phenomenology for the essential features of superconductivity in disordered systems (i) the spot to spot spatial variation on of the superconducting gap in cuprates detected by scanning tunneling microscope (STM) recording the static picture of states pinned at the dopant sites at the sample surface \cite{McElroy}, and (ii) the recent characterization of the superconductor-insulator  transition by increasing continuously the charge density by gate voltage when the normal state resistance decreases below the value of the quantum resistance for pairs $R_Q = 6.5k\Omega$ \cite{Bozovic}.

The focus is recently addressed on controlling oxygen defects  both interstitials and vacancies  to provide many avenues to control superconducting functionalities \cite{Littlewood}. In fact it has been shown recently that oxygen defects in heterostructures control  other electronic states, including magnetism and ferroelectricity \cite{brink}, a two-dimensional electron gas with universal sub-bands can arise from the oxygen vacancies confined in a surface layer at atomic limit in a cleaved SrTiO3 crystals  \cite{dagotto2,meevasana,santander} 
and it could show superconductivity \cite{reyren}. It has bee proposed that the defects clustering in cuprates might provide a form of optimal inhomogeneity for superconductivity \cite{geballe}. A fingerprint of structural scale invariance of dopants has been reported recently by Fratini {\it et al.}\cite{fratini} by detecting the structural scale invariance of dopants using scanning micro x-ray diffraction, a mixed $k$-space and real-space probe, and confirmed by Poccia {\it et al.} \cite{Poccia} using time resolved x-ray diffraction study of their nucleation and growth. The lattice complexity has been related with quantum criticality \cite{Zaanen2010}. Both the magnitude and spatial distributions of the dopants distribution show power-law behavior with an  exponential cutoff depending on the material and the doping, the unique fingerprint of scale invariance. Moreover the critical temperature $T_c$ for superconductivity in these materials increases by a factor of 2.5 from 16K to 40K with the value of the exponential cutoff changing by a factor of 3.8  \cite{fratini} suggesting that the fractal background favors the superconductive phase. The scientific interest on this new scenario is rapidly growing since the control of complex structural organization of dopants and defects in new functional materials could allow us to manipulate granular superconductors with many venues for both science and technology. 

In this work we consider a possible scenario \cite{Kresin1,Kresin2,muller}  where at temperatures higher than $T_c$ there are superconducting grains without phase coherence (that we will call the insulator phase) and a superconducting coherent phase state below $T_c$ (that we will call the superconducting phase). We propose that the RTIM in annealed heterogeneous networks sheds light on the increase of the high critical temperature for an optimal heterogeneity in this scenario. Our model will describe the  dynamical nature of networks that rapidly evolve in time by a rewiring of the links. Moreover  a key element of our analysis is to study the role of heterogeneous degree distribution, focusing in particular on scale-free degree distribution with an exponential cutoff which indicates the distance from a critical point which triggers the complexity of the electron background. This will open a theoretical roadmap to mimic dynamics occurring on a complex fractal disorder \cite{fratini}  and will help to contribute in the general understanding of how dynamical processes are affected by complex network topologies.  We observe a  rich interplay between network structure and quantum dynamical behavior in annealed complex networks. In particular we  have found  that the critical temperature for superconductivity is  modulated by the topology of the underlying networks similarly to what happens in the classical Ising model on scale-free networks where it has been shown \cite{ising} that the phase diagram depends on the power-law exponent of the degree distribution. Moreover  the critical temperature  depends on the power-law exponent on the degree distribution and on the exponential  cutoff in the degree distribution  that mimics the correlation between  dynamical granular patches in cuprates. These results provide a new perspective showing how complexity can increase  the critical temperature in an unconventional superconductor \cite{Poccia,Littlewood}.

{\it Annealed Complex Networks  -}
We consider networks of $N$ nodes  $i=1,\ldots,N$. We assign to each node a hidden variable $\theta_i$ from a $p(\theta)$ distribution indicating the expected number of neighbors of a node. 
The probability $(i,j)$ that two nodes are linked  $p_{ij}$ is given by 
\be
p_{ij}=\frac{\theta_i \theta_j }{\avg{\theta} N}
 \label{pij2}
\ee
In this ensemble the degree $k_i$ of a node $i$ is a Poisson random variable with expected degree $\overline{k_i}=\theta_i$. 
Therefore we will have 
\bea
\avg{\theta}&=&\avg{k}\nonumber \\
\avg{\theta^2}&=&{\avg{k(k-1)}}.
\eea

{\it Random Transverse Ising model on annealed complex networks -}
We consider a system of  spin variables 
$\sigma_i^z$, for $i=1,\dots,N$, defined on the nodes of a given annealed network with link probability
given by the matrix ${\bf p}$ and adjacency matrix ${\bf a}$.
The Random Traverse Ising Model is defined as in \cite{IoffeMezard1,IoffeMezard2} as
 \be 
\hat {H}=-\frac{J}{2}\sum_{ij}a_{ij}\hat{\sigma}^z_i \hat{\sigma}^z_j-\sum_i \epsilon_i \hat{\sigma}^x_i-h\sum_i \hat{\sigma}_i^z.
\label{H0}\ee
This Hamiltonian is a simplification respect to the $XY$ model Hamiltonian proposed by Ma and Lee \cite{MaFisher} to describe the superconducting-insulator phase transition but to the leading order the equation for the order parameter is the same, as widely discussed in \cite{IoffeMezard1, IoffeMezard2}.
The Hamiltonian describes the superconducting-insulator phase transition as a ferromagnetic spin $1/2$ spin system in a transverse field.
We propose to use this Hamiltonian to describe  in a granular superconductor the transition from a phase of superconducting grains with no phase coherence (called insulator for granular superconductors) that we propose to correspond with the  electronic matter in cuprates above $T_c$ to the low temperature superconducting phase with phase coherence.   
 
The spins  $\sigma_i$  in Eq. $(\ref{H0})$ indicate occupied or unoccupied states by a Cooper pair or a localized pair; the parameter $J$ indicates the couplings between neighboring spins, $\epsilon_i$  are quenched values of on-site energy  and $h$ is an external auxiliary field. To mimic the randomness of on-site energy we draw  the variables $\epsilon_i$ from a $\rho(\epsilon)$ distribution with a finite support.   Finally in  this model the superconducting phase corresponds to the existence of a  spontaneous magnetization in the $z$ direction.  
The partition function  for this problem is given by 
\be
Z=\mbox{Tr}\  e^{-\beta \hat{H}}
\ee
with the Hamiltonian given by Eq. $(\ref{H0})$
where in order to account for the dynamics nature of the annealed graph we have substituted the adjacency matrix $a_{ij}$ in $\hat{H}$ with the matrix $p_{ij}$ given by Eq. $\ref{pij2}$.
In order to evaluate the partition function we  apply the Suzuki-Trotter decomposition \cite{Mahan} in a number $N_s$ of Suzuki-Trotter slices. This formula expresses the exponential of the sum between two operators in terms of a limit of a product of  exponentials, i.e.
\bea
e^{A+B}=\lim_{N_s\to \infty} [e^{A/N_s} e^{B/N_s}]^{N_s},
\eea
 Therefore we have in the limit $N_s \to \infty$ that the partition function $Z$ can be written as 
\bea
Z=\mbox{Tr}\  e^{-\beta \hat{H}}=\mbox{Tr} \left(e^{-\beta \hat{E}/N_s}e^{-\beta \sum_i \epsilon_i\hat{\sigma}^x_i/N_s}\right)^{N_s},
\eea
where $\hat{E}$ is given by 
\bea
\hat{E}=-\frac{J}{2}\sum_{ij}p_{ij}\hat{\sigma}^z_i \hat{\sigma}_j^z-h\sum_i \hat{\sigma}_i^{z},
\eea
where $\hat{\sigma}_i^z$ are the spin operators.
In order to perform the calculation of the partition function $Z$ we consider for each spin the sequence ${\underline{\sigma_i}}=\{\sigma_i^1,\ldots \sigma_i^{N_s}\}$ where each spin $\sigma_i^{\alpha}$ represents the spin $i$ in the Suzuki-Trotter slice $\alpha$.
The partition function is then defined as
\bea
Z&=&\sum_{\{\underline{\sigma}_i\}_{i=1,\ldots, N}}\prod_{i=1}^N w(\underline{\sigma_i})e^{\frac{\beta h}{N_s}\sum_{\alpha=1}^{N_s}\sigma_i^{\alpha}}\nonumber \\
& &\times  e^{\frac{\beta J}{2 N_s \avg{\theta} N}\sum_{\alpha=1}^{N_s}\sum_{ij}\theta_i \theta_j \sigma_i^{\alpha}\sigma_j^{\alpha} }
\eea 
where we have indicated with 
\bea
w(\underline{ \sigma_i})=\prod_{\alpha} \langle \sigma_i^{\alpha}|e^{\frac{\beta \epsilon_i}{N_s}\sigma^x}|\sigma_i^{\alpha+1}\rangle.
\eea
In order to disentangle the quadratic terms we use $N_s$ Hubbard-Stratonovich transformations
\bea
Z&=&\left(\frac{\beta N\avg{\theta}}{2\pi N_S}\right)^{N_s/2}\int {\cal D}\underline{S} \exp[-\frac{N\avg{\theta}\beta J}{2N_s}\sum_{\alpha}(S^{\alpha})^2]\nonumber \\
&&\exp[N\sum_\theta p(\theta)\int d\epsilon \rho(\epsilon)\ln \mbox{Tr} \prod_{\alpha}e^{\frac{\beta}{N_s}(h+J \theta S^{\alpha})\sigma^z}e^{\frac{\beta}{N_s}\epsilon\sigma^x}], \nonumber
\eea
where ${\cal D}\underline S=\prod_{\alpha=1}^{N_s}d S^{\alpha}$.
The free energy $f=-\frac{1}{\beta}\lim_{N\to \infty } \lim_{N_s \to \infty} \frac{1}{N} \ln Z$ can be evaluated at the stationary saddle point which is cyclically invariant. Therefore we get
\bea
f&=&\mbox{inf}_{S } \frac{J \avg{\theta}}{2}S^2\nonumber \\
&&-\frac{1}{\beta}\sum_{\theta}p(\theta)\int d\epsilon \rho(\epsilon)\ln\left(2\cosh\left(\beta \sqrt{(h+JS\theta)^2+\epsilon^2}\right)\right)\nonumber
\eea
where the value of $S$ which minimizes the free energy is given by the saddle point equation
\bea
S&=&\sum_{\theta}\frac{\theta}{\avg{\theta}} p(\theta)\int d\epsilon \rho(\epsilon)\frac{JS\theta+h}{\sqrt{(JS\theta+h)^2+\epsilon^2}}\nonumber \\
&&\times\tanh(\beta \sqrt{(JS\theta+h)^2+\epsilon^2})
\label{sp}
\eea
Finally the magnetizations along the axis $x$ and $z$ can be calculated by evaluating 
\bea
m_{\theta,\epsilon}^z=\left.\frac{\mbox{Tr }\sigma_{i}^z e^{-\beta \hat{H}}}{Z}\right|_{\theta_i=\theta,\epsilon_i=\epsilon}\nonumber \\
m_{\theta,\epsilon}^x=\left.\frac{\mbox{Tr }\sigma_{i}^x e^{-\beta \hat{H}}}{Z}\right|_{\theta_i=\theta,\epsilon_i=\epsilon}.
\eea 
Performing these calculations we get
\bea 
m_{\theta,\epsilon}^z&=&\frac{JS\theta+h}{\sqrt{(JS\theta+h)^2+\epsilon^2}}\tanh(\beta \sqrt{(JS\theta+h)^2+\epsilon^2})\nonumber \\
m_{\theta,\epsilon}^x&=&\frac{\epsilon}{\sqrt{(JS\theta+h)^2+\epsilon^2}}\tanh(\beta \sqrt{(JS\theta+h)^2+\epsilon^2})\nonumber
\eea
Therefore the magnetizations $m_{\theta,\epsilon}^z$ and $m_{\theta,\epsilon}^x$ depend on the value  $\theta$ of the expected degree of the node and on the on-site energy $\epsilon$.

\begin{figure}
\begin{center}
\includegraphics[width=.6\columnwidth]{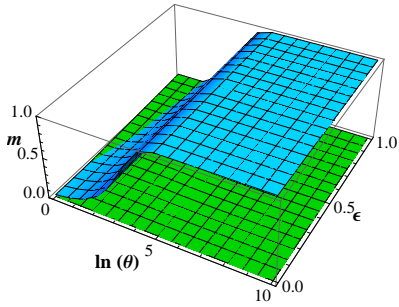}
\end{center}
\caption{(Color online)   Magnetization $m_{\theta,\epsilon}^z$ (cyan top surface) $m_{\theta,\epsilon}^x$ (green lower surface) as a function of $\theta$ and $\epsilon$ for $T=10$, expected degree distribution $p(\theta)\propto k^{-\gamma}$ and $\gamma=2.5$, and distribution of on-site energies $\rho(\epsilon)=0.5$ with $\epsilon=(-1,1)$. }
\label{due}
\end{figure}

These magnetization curves are plotted as a function of $\theta$ and $\epsilon$ in Figure $\ref{due}$ for some parameters values.
The order parameter for the superconducting-insulator phase transition is $S$ given by Eq. $(\ref{sp})$.

From the self-consistent equation determining the order parameter for the transition it is immediate to show that the  superconducting-insulator  phase transition occurs for $h=0$ at 
\be
1=J\frac{\avg{\theta^2}}{\avg{\theta} }\int d\epsilon \rho(\epsilon)\frac{\tanh(\beta \epsilon)}{\epsilon}=\frac{J}{J_c(\beta)}
\ee
which implies that for $\avg{\theta^2}\to \infty$ then $\beta\to 0$  for any fixed value of the coupling $J>0$ ,  and the critical temperature for the paramagnetic ferromagnetic phase transition $T_c$ diverges.
This implies that on annealed scale free networks with $p(\theta)\propto \theta^{-\gamma}$ and $\gamma<3$   the random  transverse Ising model  is always in the superconducting phase.

\begin{figure}
\begin{center}
\includegraphics[width=.9\columnwidth]{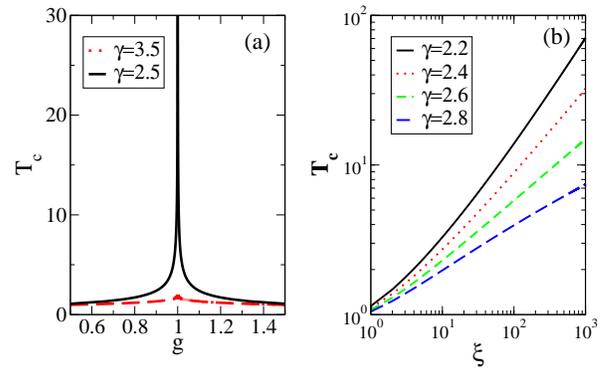}
\end{center}
\caption{(Color online)  Left panel: The critical temperature $T_c$ for the Random Transverse Ising Model with $J=1$ $\rho(\epsilon)=0.5 $ with $\epsilon\in(-1,1)$ as a functional of the external parameter $g$ tuning the exponential cutoff of the expected degree distribution $p(\theta)\propto \theta^{-\gamma}e^{-\theta/\xi}$ with $\xi=|g-g_c|^{-\nu}$,$g_c=1$ and  $\nu=1$. The critical temperature $T_c$ for the superconductor insulator transition has a maximum for $g=g_c=1$ if $\gamma>3$ and diverges at $g=g_c$ is $\gamma<3$. Right panel: critical temperature $T_c$ for the Random Tranverse Ising Model with $J=1$  $\rho(\epsilon)=0.5 $ with $\epsilon\in(-1,1)$ as a functional of the exponential cutoff in the expected degree distribution for different values of the power-law exponent $\gamma$.}
\label{uno}
\end{figure}

In order to mimic the fractal background present in cuprates \cite{fratini} we assume that the expected degree distribution of the network is given by 
\bea
p(\theta)=\theta^{-\gamma}e^{-\theta/\xi}
\eea
where $\xi $ eventually becomes much larger than the lattice coordination number and diverges as a function of an external parameter $g$ (mimicking doping of the cuprate or strain of the lattice) with a critical exponent $\nu$, i.e. we assume
\bea
\xi=|g-g_c|^{-\nu}.
\eea
In figure $\ref{uno}$ we show the behavior of the critical temperature as a function of the external parameter $g$ below and above the critical point $g_c$.
It is shown that as $\gamma>3$ the critical temperature of the RTIM has a maximum at $T_c$ while for $\gamma<3$ it diverges at $g=g_c$.

{\it Conclusions -}
In this paper we have investigated how topological effects and heterogeneity in the degree distribution  can affect the critical behavior of the superconductor-insulator transition in  annealed complex networks. In particular, we have shown how  topological effects change significantly the phase diagram of the critical phenomena. Our analysis points out that a finite  second moment of the degree distribution determines the critical lines of the transition while when this second moment diverges the critical temperature for the ferromagnetic phase diverges. In the future we plan to study how much this scenario changes if we consider quenched complex networks by applying the quantum cavity method.
Finally our results open new perspectives for the comprehension of critical phenomena  in complex materials when a structural phase transition driven by some external parameter $g$, modulates with an exponential cutoff the  scale-free heterogeneities in the material. We found that in this case two scenarios are possible: either the superconducting temperature reaches a maximum at $g=g_c$ or the superconducting temperature diverges at $g=g_c$ depending on the value of the power-law exponent $\gamma$ of the degree distribution of the annealed network. We hope that this work will be of help for the design of new complex granular high temperature superconductors that could eventually overcome the limit of 160K in cuprate superconductors.

\end{document}